\documentclass{article}
\usepackage{spconf,amsmath,graphicx,url,enumitem}

\newcommand{\eref}[1]{(\ref{#1})}

\newcommand{\fref}[1]{Fig.~\ref{#1}}
\newcommand{\tref}[1]{Table~\ref{#1}}

\title{Impulse Response Data Augmentation and Deep Neural Networks\\
for Blind Room Acoustic Parameter Estimation}
\name{Nicholas J. Bryan}
\address{Adobe Research, San Francisco, CA, USA}

\begin{document}
%
\maketitle

\begin{abstract}
The reverberation time (T60) and the direct-to-reverberant ratio (DRR) are commonly used to characterize room acoustic environments. Both parameters can be measured from an acoustic impulse response (AIR) or using blind estimation methods that perform estimation directly from speech. When neural networks are used for blind estimation, however, a large realistic dataset is needed, which is expensive and time consuming to collect. To address this, we propose an AIR augmentation method that can parametrically control the T60 and DRR, allowing us to expand a small dataset of real AIRs into a balanced dataset orders of magnitude larger. Using this method, we train a previously proposed convolutional neural network (CNN) and show we can outperform past single-channel state-of-the-art methods. We then propose a more efficient, straightforward baseline CNN that is 4-5x faster, which provides an additional improvement and is better or comparable to all previously reported single- and multi-channel state-of-the-art methods.
\end{abstract}


\section{Introduction}
\label{sec:intro}
Acoustic impulse responses (AIRs) are commonly modeled as having early- and late-field responses~\cite{schroeder1962natural, schroeder1987statistical}.  The early response consists of the direct path and early reflections imposed by the microphone-room geometry and the late-field response consists room volume and materials information. This decomposition motivates the use of the direct-to-reverberant ratio (DRR) and the reverberation time (T60) to characterize acoustic environments. The DRR is the energy ratio of sound arriving at a microphone directly from a source divided by the sound arriving after one or more surface reflections~\cite{naylor2010speech} and the T60 is the time it takes for a sound to decay 60dB within a diffuse field~\cite{kuttruff2016room}. DRR and T60 are typically measured directly from acoustic impulse responses (AIRs). In many cases, however, direct estimation is difficult, motivating blind methods that perform estimation directly from recorded speech. 
\begin{figure}[!tb]
    \centering
  \centerline{\includegraphics[width=.999\columnwidth, height=3.75cm]{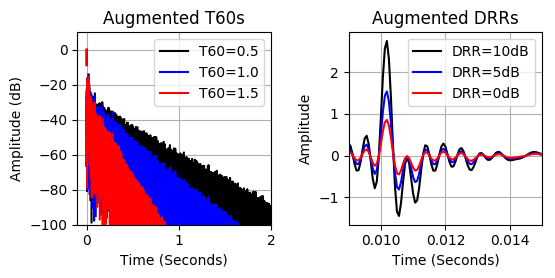}}
\kern-1em
      \caption{Acoustic impulse response augmentation. (Left) Augmentation is applied to impose varying T60 levels. (Right) Augmentation is applied to impose varying DRR levels.} \label{fig:headline}
\end{figure}

The ACE Challenge was recently held to benchmark blind DRR and T60 estimation methods~\cite{eaton2016estimation}. The best single-channel fullband DRR estimator was a machine learning (ML) approach with hand-crafted features~\cite{parada2015evaluating} and the best single-channel fullband T60 estimators were signal processing approaches~\cite{prego2015blind, loellmann2015single}. Given the recent advances in deep learning (DL), however, it is surprising that DL approaches did not outperform alternatives. When we examine further, however, we see that the ACE Challenge and other datasets are small in size, limiting the effectiveness of deep networks. 

To work around this, Xiong et al.~\cite{xiong2015joint} and Parada et al.~\cite{parada2015evaluating}, use several (five+) open-source or custom AIR datasets. Such data collection efforts, however, still result in small, unbalanced collections of AIRs, limiting performance. More recently, Gamper and Tashev (GT-CNN)~\cite{gamper2018blind} take a similar approach with the addition of using synthetic AIRs to train a compact convolutional neural network with an equivalent rectangular-bandwidth filterbank (ERB) front-end feature extractor to achieve state-of-the-art results for fullband T60 estimation. The authors, however, explicitly mention issues of small, imbalanced data, and the desire for data augmentation to improve results.
 
To address this, we propose the use of AIR augmentation with deep convolutional neural networks (CNN) to estimate T60 and DRR from speech. Data augmentation has been used to help overcome small dataset size issues in related applications~\cite{mcfee2015software, ko2017study, salamon2017deep}, but, to our knowledge, has not been applied to AIRs for blind room acoustic parameter estimation. For this, we develop a new augmentation method to parametrically control the T60 and DRR of an AIR as shown in~\fref{fig:headline}. We use the method to augment a small existing AIR dataset into a statistically balanced dataset that is orders of magnitude larger. Using this method, we adopt the GT-CNN method to both T60 and DRR estimation and show we can outperform past single-channel state-of-the-art methods significantly. We then propose a more efficient, straightforward baseline CNN that is 4-5x faster, which provides an additional improvement and suggest our baseline approach is better or comparable to all previously reported single- and multi-channel state-of-the-art methods.

\section{Impulse Response Augmentation}
\label{sec:augmentation}
To perform AIR augmentation (AIRA), we develop a procedure that allows us to parametrically control the subband DRR and T60 of a recorded AIR. Before we outline our method, we define
\begin{eqnarray}
h_e(t) &=& 
\left\{ \begin{array}{ll}
	h(t) & t_d - t_0 \leq t \leq t_d + t_0 \\
 	0 &\text{otherwise}
\end{array} \right. \label{eq:he} \\
h_l(t) &=& 
\left\{ \begin{array}{ll}
	h(t) & t < t_d - t_0   \\
	h(t) & t > t_d + t_0 \\
 	0 &\text{otherwise} \label{eq:hl},
\end{array} \right.
\end{eqnarray}
where $h(t)$ is an AIR, $t$ is a discrete time index, $h_e(t)$ is the early response, $h_l(t)$ is the late-field response, $t_d$ is the time delay of the direct path, and $t_0$ is tolerance window set to 2.5 ms~\cite{eaton2016estimation}. We identify the location of the direct path as the time of the maximum of the absolute value of $h(t)$.  


\subsection{Direct-to-reverberant ratio augmentation}
\label{sec:drr_augmentation}
\begin{figure}[t!]
\centerline{\includegraphics[width=.95\columnwidth, height=3.65cm]{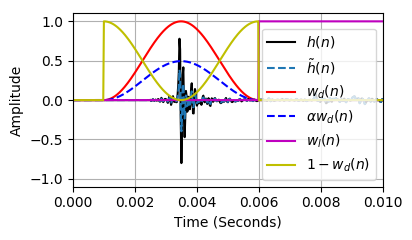}}
  \kern-.5em
\caption{Augmentation of the DRR. The direct component is windowed, scaled, and then mixed together with late-field.}
\label{fig:drr_augmentation}
\end{figure}

The DRR is defined as 
\begin{equation}
\mathrm{DRR_{dB}} = 10 \log_{10} \Bigg (  \frac{\sum_{t} h^2_e(t) }{\sum_{t}h^2_l(t)  }   \Bigg )
\label{eq:drr}.
\end{equation}
To augment the DRR, we can apply a scalar gain, $\alpha$, to the early response $h_e(t)$ via $h_e(t) \leftarrow h_e(t) \alpha$, where $\alpha$ can be chosen to fit any desired DRR. To avoid introducing discontinuities and maintain realistic AIRs, however, we further decompose the early response into a windowed direct path and windowed residual,
\begin{equation}
h_e(t) = \alpha w_d(t) h_e(t) + [1-w_d(t)] h_e(t),
\label{eq:window}
\end{equation}
as shown in~\fref{fig:drr_augmentation}, where $w_d(n)$ is set to be a 5 ms Hann window. Given a desired DRR, we rearrange~\eref{eq:window} and~\eref{eq:drr}, and solve for $\alpha$ by computing the maximum root of the quadratic equation, 
\begin{equation}
\begin{split}
 \alpha^2 \sum_{t} w^2_d(t) h^2_e(t)   + 2 \alpha \sum_t[1-w_d(t)]w_d(t)h^2_e(t) \\ + \sum_t(1-w_d(t))^2h^2_e(t) - 10^{DRR_{dB} / 10} \sum_t h^2_l(t) = 0,
 \end{split}
\label{eq:drr2}
\end{equation}
allowing us to smoothly crossfade between the manipulated early response and late-field response. To ensure that the direct path time-of-arrival does not change as a result of the scaling, we further compute the maximum of the absolute value of the late response and compare it with the original direct path maximum. If the late-field maximum is greater than the early response, we clip the applied scaling factor.  This imposes an empirical lower bound on the DRR, but in practice we did not find this to be an issue. 

\subsection{Reverberation time augmentation}
\label{sec:reverb_time_augmentation}
\begin{figure}[t]
    \centering
  \centerline{\includegraphics[width=.95\columnwidth, height=3.65cm]{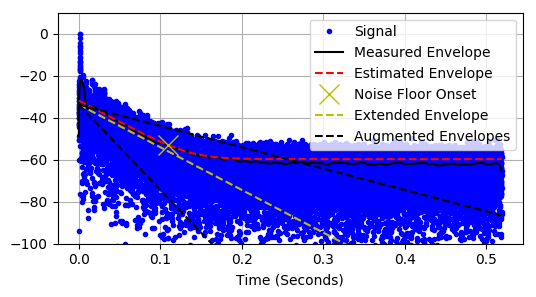}}
  \kern-.5em
      \caption{Augmentation of T60. The energy envelope of an AIR is measured (black solid), estimated (red dashed), extended (yellow dashed), and scaled to a desired level (black dashed). }      \label{fig:extended}
\end{figure}
The T60 or decay time of an AIR can be modeled and estimated in a variety of ways. Commonly, the late-field response $h_l(t)$ of an AIR is modeled as frequency-dependent exponentially decaying Gaussian noise with an additive noise floor,
\begin{equation}
h_{m}(t) = A_m e^{-(t-t_0)/\tau_m}n(t)u(t-t_0) + \sigma_m n(t), \label{eq:airmodel}
\end{equation}
where $A_m$ is the equalization level, $\tau_m$ is the decay rate, $\sigma_m$ is the noise floor level, $n(t)$ is Gaussian random noise with zero mean and unit variance,  ${T}_{60} = \ln (1000) \tau T_s$, $T_s$ is the sampling time, $t_0$ is the late-field onset time, $m$ is a frequency subband index, and $u(t)$ is a unit step response. Given the model, we estimate the decay rate, noise floor, and equalization level via the non-linear optimization method of Karjalainen (K-T60)~\cite{Karjalainen2001estimation}, which provides a parametrized two-stage energy decay envelope of an AIR.

Given estimates ($\hat{A}_m$, $\hat{\tau}_m$, $\hat{\sigma}_m$) and a desired subband decay rate $\tau_{m,d}$, we augment the subband reverberation time by multiplying~\eref{eq:airmodel} by a growing or shrinking exponential
\begin{equation}
h_m(t) \leftarrow h_m(t) e^{-(t - t_0) \frac{\hat{\tau}_m - \tau_{m,d}}{\hat{\tau}_m\tau_{m,d}}}.
\label{eq:expand}
\end{equation}
In doing so, we undo the real decay time of the impulse response and impose our own desired decay per frequency band. For a fullband augmentation variant that maintains the frequency-dependent decay shape, we 
\begin{enumerate}[itemsep=1mm]
\item Compute the ratio of a desired fullband decay $\tau_d$ over the estimated fullband decay $\gamma = \tau_d / \hat{\tau}$,
\item Compute augmented subband decay rates $\tau_{m,d} = \gamma \hat{\tau}_m$, 
\item Apply~\eref{eq:expand} to each subband, 
\item Sum each subband to create the final result.
\end{enumerate}
Before applying~\eref{eq:expand}, however, great care must be taken to avoid an unstable exponential growing late-field caused by the noise floor in~\eref{eq:airmodel}.  

To remove the noise floor, we follow a similar procedure as~\cite{irextension}, which detects and removes the noise floor and then stitches in a synthesized matching late-field response. In our work, however, we update the method to operate within the framework of the K-T60 estimator, which was found to be more robust and accurate~\cite{eaton2016estimation}. This is done by
\begin{enumerate}[itemsep=1mm]
\item Estimating the parametrized two-stage energy decay curve per frequency band via the K-T60 method,
\item Detecting the noise floor onset time via numerical search on the estimated decay curve,
\item Generating a modified energy envelope with the noise floor set to zero,
\item Synthesizing a Gaussian noise signal and imposing the corresponding noise-free energy envelope,
\item Cross-fading the original and synthesized late-field signal at the noise floor onset. 
\end{enumerate}
The modeling and augmentation process is illustrated in~\fref{fig:extended}. We perform this process per subband via a zero-phase power complementary third-octave filterbank with Butterworth prototypes~\cite{linkwitz1976active} and sum each subband to create the final result.   
 
\subsection{Example dataset augmentation} \label{sec:augmented_dataset}
To create an example dataset, we collect speech, noise, and AIRs, separately. For speech, we use the Device and Produced Speech (DAPS) dataset~\cite{mysore2015can}, which consists of 20 speakers reading public domain stories (4.5 hours). We split all speech files randomly ensuring each speaker is only represented in a single partition, segment the data into 8 second non-overlapping chunks, and normalize each chunk to -23 loudness units full scale (LUFS)~\cite{LUFSref}. This results in training (1,130), validation (388), and test (369) files. For noise, we use the first-channel of the  ground truth noise files from the ACE corpus development set (Building Lobby and Office 1), segment the noise data into eight second non-overlapping chunks, and split the files randomly into training (1,007), validation (257), and test (316) files.  
    
\begin{figure}[!tb]
    \centering
  \centerline{\includegraphics[width=.95\columnwidth, height=3.45cm]{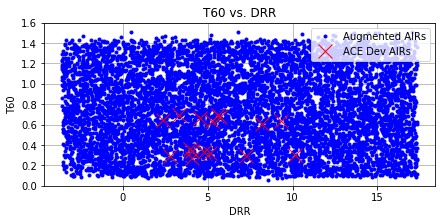}}
  \vspace{-.375cm}
      \caption{Distribution of T60 and DRR values. The augmented AIRs achieve a balanced distribution of T60 and DRR values. }\label{fig:distribution}
\end{figure}

For AIRs, we use the first channel of 16 AIRs provided by the ACE corpus development set (Building Lobby and Office 1). We apply the fullband variant of our T60 and DRR augmentation procedure in sequence 500 times per AIR. We specify T60 values to be uniformly distributed between .1-1.5 seconds and DRR values to be uniformly distributed between -6-18dB, resulting in 8000 AIRs with a balanced distribution as shown in~\fref{fig:distribution}. We then split all AIRs into training (5,120), validation (1,280), and test (1,600) files. 
\label{sec:example_augmentation}

To calibrate the our T60 and DRR ground truth estimators used label our dataset, we use linear regression (slope and intercept) to match the ACE corpus development set labels. We do this because we do not have access to the ground truth ACE estimators. We hypothesize that a lack of calibration, in addition large training data, has significantly contributed to the lack of performance of DL methods for blind parameter estimation. In~\tref{table:truth}, we show a comparison of the mean squared error (MSE), bias, and Pearson correlation coefficient of our calibrated ground truth estimator implementations compared to the ACE corpus labels. 
\begin{table}[!t]
\begin{center}
    \begin{tabular}{ | l | l | l |  p{1cm} |}
    \hline
     \textbf{Method} 					& \textbf{Bias} 		& \textbf{MSE} 	&  	$\rho$ 	  	\\ \hline
      T60	 			& 0.000 		& 0.003  	& 	0.931		\\ \hline
      DRR	 			& 0.000 		& 1.311  	& 	0.947		\\ \hline
    \end{tabular}
\end{center}
\vspace{-.35cm}
     \caption{MSE, bias, and Pearson correlation between our calibrated ground truth estimators vs. ACE AIR labeled data.} \label{table:truth}
\end{table}

Given this data, create mixture training, validation, and testing datasets. For each partition, we take each speech recording and convolve it with a random AIR. We sample random noise, circularly shift it by a random amount, randomly scale the noise to impose  uniformly distributed SNR between 20 and -5 dB, and add it to the convolved speech. For SNR, we use the ITU-T P.56 specification~\cite{itup56} for speech and a root-mean square (RMS) estimator for noise. Once mixed, we randomly sample a four second segment to produce a final sample, re-selecting any segment with an RMS level below 20dB the full-length segment. We repeat this for each speech segment in each partition 100x to create (113,000) training, (38,800) validation, and (36,900) test files.

\section{Blind Acoustic Parameter Estimation}
\label{sec:network}
We train two separate CNNs with identical preprocessing, network architecture, cost function, and training procedure. 

\subsection{Preprocessing}
\label{sec:preprocess}
We convert our mixture data into a decibel-scaled Melspectrogram representation~\cite{mcfee2015librosa} with a fast Fourier transform (FFT) and Hann window size of 256 samples (16ms), hop size of 128 samples (8ms), 32 Mel-frequency bands with area norm, and sampling rate of 16kHz, resulting in a 32 x 499 feature representation (2x FFT/hop size, 2x Mel bands, and/or data normalization had little effect on estimation results).

\subsection{Network architecture \& training}
\label{sec:arch}
Our CNN architecture for both T60 and DRR includes six 2D convolutional (conv) layers, each followed by a rectified linear activation function, max pooling, and batch normalization.  The first two conv layers have 8 kernels with size 1x2. The third and fourth conv layers have 16 kernels have with size 1x2. The fourth and fifth conv layers have 32 kernels with size 2x2. The first four conv/pooling layers reduce the dimension of the time-axis only until the time and frequency axes are approximately the same dimension. The last two conv/pooling layers reduce the dimension over both time and frequency. After the conv layers, a dropout layer (50\%) and fully connected layer are used to predict a scalar value. The max pooling size is identical to the conv layer filter size for each layer, respectively. The network has 8,737 trainable parameters and 224 non-trainable parameters. 

We train our networks to minimize the mean square error (MSE) using the Adam optimizer~\cite{kingma2014adam} with 0.01 learning rate, learning rate reduction (.5x) on plateau and early stopping with a batch size of 128. The model with the lowest validation error was selected for evaluation.  For inference, we slide our estimator over time and compute an estimate every half second. We use the median of all estimates for a stationary recording. For recordings shorter than four seconds, we repeat the example until it is greater than four seconds. 

\section{Evaluation}
\label{sec:eval}
We compare our proposed CNN network trained on our data (Our CNN + AIRA) with previously published state-of-the-art results from the ACE challenge~\cite{eaton2016estimation}, previously published GT-CNN results~\cite{gamper2018blind} for T60 only, and our reimplementation of the GT-CNN~\cite{gamper2018blind} estimator trained on our augmented dataset (GT-CNN + AIRA) for both T60 and DRR.

\subsection{ACE evaluation}
\begin{table}[!tb]
\begin{center}
    \begin{tabular}{ | l | l | l |  p{1cm} |}
    \hline
     \textbf{Method} 					& \textbf{Bias} 		& \textbf{MSE} 		&  	$\boldsymbol{\rho}$ 	  	\\ \hline
     MLP~\cite{xiong2015joint}	 		            & -0.0967 			& 0.104  			& 	0.48							\\ \hline
     QA Reverb~\cite{prego2015blind}	 	        & -0.068 			& 0.0648  			& 	0.778						\\ \hline
     GT-CNN~\cite{gamper2018blind}	 	        & 0.0304 			& 0.0384  			& 	0.836						    \\ \hline
     GT-CNN~\cite{gamper2018blind}$^*$ + AIRA	 	& \textbf{0.0221} 			& 0.0265  			& 	0.9089						    \\ \hline
     Our CNN + AIRA		                & {-0.0264} 	& \textbf{0.0261}  	& 	\textbf{0.9197}				    \\ \hline
    \end{tabular}
\end{center}
\vspace{-.45cm}
     \caption{Blind T60 estimation results.} \label{table:t60results}
\end{table}
\begin{table}[!tb]
\begin{center}
    \begin{tabular}{ | l | l | l |  p{1cm} |}
    \hline
     \textbf{Method} 							 & \textbf{Bias} 		& \textbf{MSE} 	&  	$\boldsymbol{\rho}$ 	  	\\ \hline
     PSD beamspace+bias$^*$~\cite{hioka2015psd} 		 & 1.07$^*$			& \textbf{8.14}$^*$		& 	0.577$^*$		\\ \hline
     NIRAv2~\cite{parada2015evaluating}			  & -1.85			& 14.8			& 	0.558		\\ \hline
     GT-CNN~\cite{gamper2018blind}$^*$ + AIRA & 1.3141          & 10.6316       &   0.6818  \\ \hline
     Our CNN + AIRA	 					  & \textbf{0.8075}	& \textbf{8.9783}  			& \textbf{0.7077}		\\ \hline
    \end{tabular}
\end{center}
\vspace{-.45cm}
     \caption{DRR estimation results. $^*$ denotes multi-channel.} \label{table:drrresults}
\end{table}

\tref{table:t60results} and \tref{table:drrresults} show T60 and DRR estimation bias, mean squared error, and Pearson correlation coefficient $\rho$ results. Compared to the previously published GT-CNN~\cite{gamper2018blind} results, we see that using AIRA data outperforms the prior state-of-the-art with a relative improvement of +27\%, +31\%, +8\% for bias, MSE, and correlation coefficient, respectively. Using our CNN + AIRA, the improvement is 13\%, 32\%, 10\%.

For DRR, when we adopt the GT-CNN~\cite{gamper2018blind} method to DRR and use AIRA (GT-CNN$^*$ + AIRA), we outperform the past single-channel state-of-the-art DRR estimation method of NIRAv2~\cite{parada2015evaluating} in terms of bias, MSE, and correlation with a relative improvement of +29\%, +28\%, +22\%. Using our CNN + AIRA, we achieve a better relative improvement of 56\% 39\% 27\%, respectively. We also outperform the state-of-the-art multi-channel PSD beamspace+bias method~\cite{hioka2015psd} in terms of bias and correlation, with comparable MSE.   

In terms of computational speed, the real-time factor (RTF) of our method is 0.0088 or over 110x real-time for T60 and DRR (independently) using a 2018 Macbook Pro with CPU-only computation. Compared to the GT-CNN method, our method is about 5x faster compared to previously published results (using different machines) and 4.74x using our implementation on the same machine. Compared to the previously report RTF for the NIRAv2 DRR method (0.899), our method is 100x faster (using different machines).
\section{Conclusions}
\label{sec:conclusions}
We propose an AIR augmentation method to control the DRR and T60 from an existing AIR. This allows us to use a small set of existing AIRs to generate a realistic, statistically balanced dataset that is orders of magnitude larger. We further propose a basic CNN for blind room acoustic parameter estimation and then compare our CNN against several baselines using the ACE corpus software. Results suggests our complete method (CNN + AIRA) outperforms past single- and multi-channel state-of-the-art T60 and DRR algorithms in terms of the correlation coefficient and bias, are either better or comparable in terms of MSE, and is at least 4-5x faster.
\bibliographystyle{IEEEbib}
\bibliography{strings,refs,refs15}

\begin{thebibliography}{10}

\bibitem{schroeder1962natural}
Manfred~R. Schroeder,
\newblock ``Natural sounding artificial reverberation,''
\newblock {\em Journal of the Audio Engineering Society}, vol. 10, no. 3, 1962.

\bibitem{schroeder1987statistical}
Manfred~R. Schroeder,
\newblock ``Statistical parameters of the frequency response curves of large
  rooms,''
\newblock {\em Journal of the Audio Engineering Society}, vol. 35, no. 5, 1987.

\bibitem{naylor2010speech}
Patrick~A Naylor and Nikolay~D Gaubitch,
\newblock {\em Speech Dereverberation},
\newblock Springer Science \& Business Media, 2010.

\bibitem{kuttruff2016room}
Heinrich Kuttruff,
\newblock {\em {Room Acoustics}},
\newblock Taylor \& Francis Group, London, U. K., 6th edition, 2016.

\bibitem{eaton2016estimation}
James Eaton, Nikolay~D. Gaubitch, Alastair~H. Moore, and Patrick~A. Naylor,
\newblock ``Estimation of room acoustic parameters: The {ACE} challenge,''
\newblock {\em IEEE/ACM Transactions on Audio, Speech and Language Processing
  (TASLP)}, vol. 24, no. 10, 2016.

\bibitem{parada2015evaluating}
Pablo~Peso Parada, Dushyant Sharma, Toon van Waterschoot, and Patrick~A.
  Naylor,
\newblock ``Evaluating the non-intrusive room acoustics algorithm with the ace
  challenge,''
\newblock in {\em ACE Challenge Workshop, A Satellite Event of Workshop on
  Applications of Signal Processing to Audio and Acoustics (WASPAA)}. IEEE,
  2015.

\bibitem{prego2015blind}
Thiago de~M. Prego, Amaro~A. de~Lima, Rafael Zambrano-L{\'o}pez, and Sergio~L.
  Netto,
\newblock ``Blind estimators for reverberation time and direct-to-reverberant
  energy ratio using subband speech decomposition,''
\newblock in {\em Applications of Signal Processing to Audio and Acoustics
  (WASPAA), 2015 IEEE Workshop on}. IEEE, 2015.

\bibitem{loellmann2015single}
Heinrich L{\"o}llmann, Andreas Brendel, Peter Vary, and Walter Kellermann,
\newblock ``Single-channel maximum-likelihood t60 estimation exploiting subband
  information,''
\newblock in {\em ACE Challenge Workshop, A Satellite Event of Workshop on
  Applications of Signal Processing to Audio and Acoustics (WASPAA)}.
  arxiv.org, 2015.

\bibitem{xiong2015joint}
Feifei Xiong, Stefan Goetze, and Bernd~T. Meyer,
\newblock ``Joint estimation of reverberation time and direct-to-reverberation
  ratio from speech using auditory-inspired features,''
\newblock in {\em ACE Challenge Workshop, A Satellite Event of Workshop on
  Applications of Signal Processing to Audio and Acoustics (WASPAA)}.
  arxiv.org, 2015.

\bibitem{gamper2018blind}
Hannes Gamper and Ivan~J. Tashev,
\newblock ``Blind reverberation time estimation using a convolutional neural
  network,''
\newblock in {\em International Workshop on Acoustic Signal Enhancement
  (IWAENC)}. IEEE, 2018.

\bibitem{mcfee2015software}
Brian McFee, Eric~J. Humphrey, and Juan~Pablo Bello,
\newblock ``A software framework for musical data augmentation.,''
\newblock in {\em ISMIR}, 2015, pp. 248--254.

\bibitem{ko2017study}
Tom Ko, Vijayaditya Peddinti, Daniel Povey, Michael~L. Seltzer, and Sanjeev
  Khudanpur,
\newblock ``A study on data augmentation of reverberant speech for robust
  speech recognition,''
\newblock in {\em International Conference on Acoustics, Speech and Signal
  Processing (ICASSP)}. 2017, pp. 5220--5224, IEEE.

\bibitem{salamon2017deep}
Justin Salamon and Juan~Pablo Bello,
\newblock ``Deep convolutional neural networks and data augmentation for
  environmental sound classification,''
\newblock {\em IEEE Signal Processing Letters}, vol. 24, no. 3, pp. 279--283,
  2017.

\bibitem{Karjalainen2001estimation}
Matti Karjalainen, Poju Antsalo, Aki Makivirta, Timo Peltonen, and Vesa
  Valimaki,
\newblock ``Estimation of modal decay parameters from noisy response
  measurements,''
\newblock in {\em Audio Engineering Society Convention 110}. Audio Engineering
  Society, 2001.

\bibitem{irextension}
Nicholas~J. Bryan and Jonathan~S. Abel,
\newblock ``Methods for extending room impulse responses beyond their noise
  floor,''
\newblock in {\em Audio Engineering Society Convention 129}. Audio Engineering
  Society, 2010.

\bibitem{linkwitz1976active}
Siegfried~H. Linkwitz,
\newblock ``Active crossover networks for noncoincident drivers,''
\newblock {\em Journal of the Audio Engineering Society}, vol. 24, no. 1, pp.
  2--8, 1976.

\bibitem{mysore2015can}
Gautham~J. Mysore,
\newblock ``Can we automatically transform speech recorded on common consumer
  devices in real-world environments into professional production quality
  speech?—a dataset, insights, and challenges,''
\newblock {\em IEEE Signal Processing Letters}, vol. 22, no. 8, 2015.

\bibitem{LUFSref}
``{Loudness Recommendation, European Broadcasting Union (EBU) Recommendation
  R128-2014},'' 2014.

\bibitem{itup56}
``{Objective Measurement of Active Speech Level, International
  Telecommunications Union (ITU-T) Recommendation P.56},'' March 1993.

\bibitem{mcfee2015librosa}
Brian McFee, Colin Raffel, Dawen Liang, Daniel~PW Ellis, Matt McVicar, Eric
  Battenberg, and Oriol Nieto,
\newblock ``librosa: Audio and music signal analysis in python,''
\newblock in {\em Proceedings of Python in Science Conference}, 2015.

\bibitem{kingma2014adam}
Diederik~P. Kingma and Jimmy Ba,
\newblock ``Adam: A method for stochastic optimization,''
\newblock {\em arXiv preprint arXiv:1412.6980}, 2014.

\bibitem{hioka2015psd}
Yusuke Hioka and Kenta Niwa,
\newblock ``{PSD} estimation in beamspace for estimating direct-to-reverberant
  ratio from a reverberant speech signal,''
\newblock in {\em ACE Challenge Workshop, A Satellite Event of Workshop on
  Applications of Signal Processing to Audio and Acoustics (WASPAA)}.
  arxiv.org, 2015.

\end{thebibliography}

\end{document}